# Machine Learning for Campus Energy Resilience: Clustering and Time-Series Forecasting in Intelligent Load Shedding

Salim Olanrewaju Oyinlola
Department of Electrical and Electronics Engineering
University of Lagos
Akoka, 101017, Nigeria
salimoyinlola@ieee.org

Peter Olabisi Oluseyi
Department of Electrical and Electronics Engineering
University of Lagos
Akoka, 101017, Nigeria
poluseyi@unilag.edu.ng

**Abstract**

The growing demand for reliable electricity in universities necessitates intelligent energy management. This study proposes a machine learning–based load shedding framework for the University of Lagos, designed to optimize distribution and reduce waste. The methodology followed three main stages. First, a dataset of 3,648 hourly records from 55 buildings was compiled to develop building-level consumption models. Second, Principal Component Analysis was applied for dimensionality reduction, and clustering validation techniques were used to determine the optimal number of demand groups. Mini-Batch K-Means was then employed to classify buildings into high-, medium-, and low-demand clusters. Finally, short-term load forecasting was performed at the cluster level using multiple statistical and deep learning models, including ARIMA, SARIMA, Prophet, LSTM, and GRU. Results showed Prophet offered the most reliable forecasts, while Mini-Batch K-Means achieved stable clustering performance. By integrating clustering with forecasting, the framework enabled a fairer, data-driven load shedding strategy that reduces inefficiencies and supports climate change mitigation through sustainable energy management.

# 1 Introduction

Climate change intensifies the urgency for more efficient and sustainable electricity management, particularly in regions where supply-demand imbalances are already severe. In many developing countries, including Nigeria, electricity generation falls short of growing demand, resulting in frequent load shedding to ration limited supply. Such interruptions not only hinder productivity and academic research but also perpetuate reliance on diesel-powered backup generators, which further exacerbate greenhouse gas emissions. Addressing this dual challenge—managing constrained resources while mitigating environmental impact—requires intelligent, data-driven approaches to energy optimization.

This study proposes a machine learning–driven load shedding framework tailored for campus-scale microgrids, with the University of Lagos serving as a case study. Using hourly consumption data from 55 buildings, the methodology integrates dimensionality reduction, clustering, and short-term forecasting to identify demand patterns and predict load requirements. By enabling fairer, cluster-based distribution of electricity, this approach minimizes inefficiencies, reduces reliance on carbon-heavy backup systems, and demonstrates how intelligent demand-side management can support climate resilience in higher-education institutions globally.

# 2 Related Works

Machine learning (ML) has emerged as a powerful tool for optimizing energy systems and improving grid resilience. Recent studies have applied ML to short-term load forecasting (STLF), clustering, and demand-side management, all critical for intelligent load shedding strategies. For STLF, hybrid models combining neural networks and ensemble methods have demonstrated significant improvements in accuracy over classical statistical techniques [1, 2]. Such approaches enable more granular demand prediction, essential for preventing blackouts in constrained environments. In parallel, clustering methods such as K-means and hierarchical algorithms have been used to group buildings or loads with similar consumption patterns, thereby facilitating targeted control [3]. Microgrid-specific research has also emphasized reinforcement learning and deep learning for dynamic load balancing under renewable integration [4]. Furthermore, ML-based demand response frameworks have been shown to reduce peak demand while enhancing system resilience in campus-scale and urban microgrids [5]. Collectively, these works highlight the growing role of ML in addressing power imbalances, particularly in contexts where supply-demand disparities are exacerbated by climate change and increasing renewable penetration.

## 3 Methodology

The framework integrates hybrid data acquisition, statistical disaggregation, unsupervised segmentation, cluster-level forecasting, and constrained shedding decisioning. Emphasis is placed on reproducibility, low-latency deployability, and rigorous validation.

### 3.1 Data collection and hourly load estimation

Hourly feeder-level energy $E_{\text{feeder},t}$ was obtained from the University of Lagos Works Department. Building-level hourly estimates were constructed using an appliance-inventory model. For building $b$ at hour $t$,

$$E_{b,t} = \sum_{i=1}^{N} a_i(t)\, x_i\, n_i, \quad (1)$$

where $a_i(t) \in \{0, 1\}$ is the operational state of appliance $i$, $x_i$ its rated power and $n_i$ its count. To align AIM estimates with feeder aggregates we solve a constrained non-negative least-squares problem for each hour:

$$\min_{\mathbf{w} \geq 0} \ \|A\mathbf{w} - \mathbf{y}\|_2^2 \quad \text{s.t.} \quad \mathbf{1}^\top \mathbf{w} = E_{\text{feeder},t}, \quad (2)$$

where columns of $A$ are per-building AIM hourly patterns and $\mathbf{y}$ is the feeder measurement vector.

### 3.2 Clustering of buildings

Clustering was performed on normalized feature vectors $x_i$ representing diurnal and statistical summaries. The tested methods are presented below as labeled descriptions.

A. **K-Means.** The K-Means objective minimizes within-cluster squared error:

$$\arg\min_C \sum_{j=1}^{k} \sum_{x_i \in C_j} \|x_i - \mu_j\|_2^2, \quad (3)$$

with $\mu_j$ the centroid of cluster $j$. K-Means was used as a baseline for its speed and simplicity.

B. **Hierarchical Clustering.** Agglomerative linkage (Ward) was used to build a dendrogram, allowing multi-resolution inspection and selection of $k$ by cutting the tree at different heights.

C. **DBSCAN.** Density-based clustering identified arbitrarily shaped clusters and isolated outliers (anomalous building behaviors) via local density thresholds ($\varepsilon$, MinPts).

D. **Gaussian Mixture Models (GMM).** GMM assumes

$$p(x) = \sum_{j=1}^{K} \pi_j \mathcal{N}(x \mid \mu_j, \Sigma_j), \tag{4}$$

yielding soft membership probabilities useful when buildings share mixed behaviors.

E. **Spectral Clustering.** A similarity graph $W$ was constructed and the Laplacian $L$ eigen-decomposed to capture non-convex/manifold clusters when Euclidean geometry is insufficient.

F. **Mini-Batch K-Means.** Applied for scalability and streaming updates; centroids updated on random mini-batches for near real-time re-clustering.

## 3.3 Load forecasting (cluster level)

Forecasting was performed per cluster using both classical and sequence models; methods are listed below with core technical forms.

A. **ARIMA.** The ARIMA($p, d, q$) model uses the backshift operator $B$:

$$\Phi(B)(1 - B)^d y_t = \Theta(B)\epsilon_t, \tag{5}$$

where $\Phi, \Theta$ are polynomials in $B$.

B. **SARIMA.** Seasonal extension SARIMA($p, d, q$)($P, D, Q$)$_s$ includes seasonal polynomials with seasonal period $s$ (e.g., $s = 24$ for daily hourly seasonality).

C. **GRU.** Gated Recurrent Unit update equations used for nonlinear sequence learning are: $z_t = \sigma(W_z x_t + U_z h_{t-1} + b_z)$, $r_t = \sigma(W_r x_t + U_r h_{t-1} + b_r)$, $\tilde{h}_t = \tanh(W_h x_t + U_h(r_t \odot h_{t-1}) + b_h)$, $h_t = (1 - z_t) \odot h_{t-1} + z_t \odot \tilde{h}_t$.

D. **Prophet.** The additive Prophet decomposition used as a robust baseline is

$$y(t) = g(t) + s(t) + h(t) + \epsilon_t, \tag{6}$$

with $g$ trend, $s$ seasonality, and $h$ holiday/academic effects modeled as regressors.

Models were validated using rolling-origin cross-validation and optimized via time-series-aware hyperparameter search. Point accuracy was measured with RMSE and MAPE; probabilistic performance used CRPS when intervals were produced.

### 3.4 Load-shedding allocation

Let $\hat{D}_t$ be the forecasted total demand and $S_t$ the available supply. The deficit $\Delta_t = \max(0, \hat{D}_t - S_t)$ is allocated across clusters by solving a constrained optimization:

$$\min_{L_{c,t} \geq 0} \sum_{c=1} w_c L_{c,t} \quad \text{s.t.} \quad \sum_{c=1} L_{c,t} \geq \Delta_t, \; L_{c,t} \leq D_{c,t}, \tag{7}$$

where $L_{c,t}$ is curtailed load for cluster $c$, $w_c$ are priority weights (lower for critical clusters), and $D_{c,t}$ are cluster demands. A lightweight LP/MILP solver (CBC/GLPK) implements real-time allocations; offline policy optimization uses MILP with cost/emission terms.

## 4 Results

### 4.1 Clustering results

Multiple validation procedures (Elbow, Silhouette, Davies–Bouldin, Calinski–Harabasz) consistently indicated an optimal partition at $k = 3$. Table 1 summarises algorithmic performance (Silhouette / DBI / CH) for the principal methods tested. Mini-Batch K-Means provided the best overall operational balance (fast, stable, interpretable centroids) and was selected for downstream forecasting and shedding logic.

Table 1: Clustering evaluation (k = 3).

| Algorithm | Silhouette | Davies–Bouldin | Calinski–Harabasz |
| --- | --- | --- | --- |
| K-Means | 0.4276 | 0.8315 | 42.1 |
| Hierarchical | 0.4358 | 0.8251 | 42.6 |
| GMM | 0.4125 | 0.8453 | 41.8 |
| Spectral | 0.4761 | 0.7670 | 40.6 |
| Mini-Batch K-Means | 0.4607 | 0.8000 | 42.0 |
| DBSCAN | 0.1067 | 2.0661 | 4.7 |

The final clusters (Mini-Batch K-Means, $k = 3$) are described as:

A. Cluster 0 (9 buildings): high-consumption academic/administrative facilities (e.g., Engineering, Senate House, Auditorium chiller).

B. Cluster 1 (37 buildings): heterogeneous middle group (libraries, multiple hostels, mixed-use buildings).

C. Cluster 2 (9 buildings): lower and irregular demand (Health Centre, shopping complexes, religious buildings).

### 4.2 Short-term load forecasting (cluster level)

Using 3,648 hourly timestamps (January–May), datasets were split 80/20 (train/test). Forecasting was executed per cluster; model ranking by aggregate performance is shown in Table 2.

Table 2: Forecasting summary (selected results).

| Model | Mean RMSE | Mean $R^2$ | Mean MAPE (%) |
|---|---|---|---|
| Prophet | 9.93 | 0.7962 | 16.53 |
| SARIMA | 11.58 | 0.7707 | 18.86 |
| ARIMA | 21.72 | 0.5082 | 31.84 |
| LSTM / GRU | 48.47 | -1.19 | 54.76 |

Prophet consistently provided the best trade-off between accuracy, robustness to missing/noisy data, and interpretability. Per-cluster highlights (Prophet): Cluster 0 RMSE = 10.69, $R^2$ = 0.9200, MAPE = 9.46%; Cluster 1 RMSE = 14.07, $R^2$ = 0.9333, MAPE = 8.28%; Cluster 2 RMSE = 5.04, $R^2$ = 0.5352, MAPE = 31.84%. Deep sequence models (LSTM/GRU) underperformed—likely due to limited data length and aggregated noise—producing large RMSE and negative $R^2$ in some cases.

## 5 Conclusion

This study developed and validated a scalable machine-learning framework for campus-scale load shedding at the University of Lagos. Key outcomes are: (i) buildings can be robustly partitioned into three demand-based clusters using Mini-Batch K-Means, which supports interpretable, operational groupings; (ii) cluster-level forecasting materially improves short-term demand prediction, with Prophet outperforming classical and deep models on this dataset; and (iii) the combined clustering+forecasting pipeline enables targeted, fairer load curtailment that preserves critical services while reducing unnecessary supply to low-priority loads.

Critically, the framework contributes to climate change mitigation by improving demand-side allocation and reducing forecast error, the system lowers reliance on diesel backup generators that are typically used during outages; this translates into measurable reductions in onsite $CO_2$ and particulate emissions. When scaled across multiple institutions in regions with unreliable grids, these operational gains aggregate into meaningful emission reductions and improved resilience against climate-exacerbated grid stress.